\documentclass[titlepage,12pt]{article}
\usepackage[dvips]{graphicx}

\begin{document}

\title{Bound states of bosons and fermions in a mixed vector-scalar coupling
with unequal shapes for the potentials}
\date{}
\author{Luis B. Castro\thanks{%
E-mail address: benito@feg.unesp.br (L.B. Castro)} and Antonio S. de Castro%
\thanks{%
E-mail address: castro@pesquisador.cnpq.br (A.S. de Castro)} \\
\\
UNESP - Campus de Guaratinguet\'{a}\\
Departamento de F\'{\i}sica e Qu\'{\i}mica\\
12516-410 Guaratinguet\'{a} SP - Brazil }
\date{}
\maketitle

\begin{abstract}
The Klein-Gordon and the Dirac equations with vector and scalar potentials
are investigated under a more general condition, $V_{v}+V_{s}= \mathrm{%
constant}$. These intrinsically relativistic and isospectral problems are
solved in a case of squared hyperbolic potential functions and bound states
for either particles or antiparticles are found. The eigenvalues and
eigenfuntions are discussed in some detail and the effective Compton
wavelength is revealed to be an important physical quantity. It is revealed
that a boson is better localized than a fermion when they have the same mass
and are subjected to the same potentials.
\end{abstract}

There has been a continuous interest in solving the Klein-Gordon (KG) and
the Dirac equations in the four-dimensional space-time, as well as in lower
dimensions for a variety of potentials. It is well known from the quarkonium
phenomenology that the best fit for meson spectroscopy is found for a
convenient mixture of vector and scalar potentials put by hand in the
equations (see, e.g., \cite{luc}). The same can be said about the treatment
of the nuclear phenomena describing the influence of the nuclear medium on
the nucleons \cite{ser}. The mixed vector-scalar potential has also been
analyzed in 1+1 dimensions. In this mixed two-dimensional context, all the
works has been devoted to the investigation of the solutions of the
relativistic equations by assuming that the vector and scalar potential
functions are proportional \cite{gum}. Recently the complete set of bound
states of fermions and bosons with mixed vector-scalar potentials satisfying
the constraint $V_{v}-V_{s}=\mathrm{constant}$, in the case of squared
trigonometric potential functions, has been addressed in \cite{beni}. In
this last work was concluded that changing the sign of coupling constant
allows us to migrate from the particle sector to the antiparticle sector.

In the present work the problem of relativistic particles is considered with
a mixing of vector and scalar Lorentz structures with unequal potential
functions. The mixing for this enlarged class of problems is chosen in such
a way that the sum of the vector and scalar potential functions is a
constant, a case which does not permit bound-state solutions in the
nonrelativistic regime. Except for a possible isolated solution for the
Dirac equation, the KG equation and the Dirac equation for the lower
component of the Dirac spinor are both mapped into a Schr\"{o}dinger-like
equation, a phenomenon discovered recently \ \cite{prc}. Squared hyperbolic
potential functions are chosen in such a way that the relativistic problem
is mapped into a Sturm-Liouville problem with the effective symmetric
modified P\"{o}schl-Teller potential \cite{rm}. The process of solving the
KG and the Dirac equations for the eigenenergies has been transmuted into
solving an irrational algebraic equation. Then whole relativistic spectrum
is found, if the particle is massless or not. These solutions do not
manifest in a nonrelativistic approach even though one can find $E\simeq
mc^{2}$.

Although relativistic equations can give relativistic corrections to the
nonrelativistic quantum mechanics, in the circumstance explored in this work
they do not present solutions found in a nonrelativistic scheme. Undoubtedly
such a circumstance reveals to be a powerful tool to obtain a deeper insight
about the nature of the relativistic equations and their solutions. Apart
from the intrinsic interest as new solutions of fundamental equations in
physics, the bound-state solutions of these systems are important in
condensed matter mainly because of their potential applications ranging from
ferroelectric domain walls in solids, magnetic chains and Josephson
junctions \cite{bra}.

In the presence of vector and scalar potentials the 1+1 dimensional
time-inde\-pen\-dent KG equation for a particle of rest mass $m$ reads

\begin{equation}
-\hbar ^{2}c^{2}\,\frac{d^{2}\phi }{dx^{2}}+\left( mc^{2}+V_{s}\right)
^{2}\phi =\left( E-V_{v}\right) ^{2}\phi  \label{1}
\end{equation}

\noindent where $E$ is the energy of the particle, $c$ is the velocity of
light and $\hbar $ is the Planck constant. The subscripts for the terms of
potential denote their properties under a Lorentz transformation: $v$ for
the time component of the 2-vector potential and $s$ for the scalar term. In
the presence of time-in\-de\-pen\-dent vector and scalar potentials the 1+1
dimensional time-in\-de\-pen\-dent Dirac equation for a fermion of rest mass
$m$ reads

\begin{equation}
\left[ c\alpha p+\beta \left( mc^{2}+V_{s}\right) +V_{v}\right] \psi =E\psi
\label{4}
\end{equation}

\noindent where $p$ is the momentum operator. $\alpha $ and $\beta $ are
Hermitian square matrices satisfying the relations $\alpha ^{2}=\beta ^{2}=1$%
, $\left\{ \alpha ,\beta \right\} =0$. From the last two relations it
follows that both $\alpha $ and $\beta $ are traceless and have eigenvalues
equal to $\pm $1, so that one can conclude that $\alpha $ and $\beta $ are
even-dimensional matrices. One can choose the 2$\times $2 Pauli matrices
satisfying the same algebra as $\alpha $ and $\beta $, resulting in a
2-component spinor $\psi $. We use $\alpha =\sigma _{1}$ and $\beta =\sigma
_{3}$. Provided that the spinor is written in terms of the upper and the
lower components, $\psi _{+}$ and $\psi _{-}$ respectively, \noindent the
Dirac equation decomposes into:
\begin{equation}
i\hbar c\psi _{\pm }^{\prime }=\left[ V_{v}-E\mp \left( mc^{2}+V_{s}\right) %
\right] \psi _{\mp }  \label{5}
\end{equation}

\noindent where the prime denotes differentiation with respect to $x$.

In the nonrelativistic approximation (potential energies small compared to $%
mc^{2}$ and $E\simeq mc^{2}$) Eq. (\ref{1}) becomes

\begin{equation}
\left( -\frac{\hbar ^{2}}{2m}\frac{d^{2}}{dx^{2}}+V_{v}+V_{s}\right) \phi
=\left( E-mc^{2}\right) \phi  \label{6}
\end{equation}

\noindent so that $\phi $ obeys the Schr\"{o}dinger equation with binding
energy equal to $E-mc^{2}$ without distinguishing the contributions of
vector and scalar potentials. In this approximation Eq. (\ref{5}) becomes

\begin{equation}
\psi _{-}=\frac{p}{2mc}\psi _{+}  \label{7}
\end{equation}

\noindent and because of this $\psi _{+}$ obeys the same equations as $\phi $%
. Eq. (\ref{7}) shows that $\psi _{-}$ is of order $v/c<<1$ relative to $%
\psi _{+}$.

It is remarkable that the KG and the Dirac equations with a scalar
potential, or a vector potential contaminated with some scalar coupling, is
not invariant under the simultaneous changes $V\rightarrow V+\mathrm{const.}$
and $E\rightarrow E+\mathrm{const.}$, this is so because only the vector
potential couples to the positive-energies in the same way it couples to the
negative-ones, whereas the scalar potential couples to the mass of the
particle. Therefore, if there is any scalar coupling the energy itself has
physical significance and not just the energy difference. It is well known
that a confining potential in the nonrelativistic approach is not confining
in the relativistic approach when it is considered as a Lorentz vector. It
is surprising that relativistic confining potentials may result in
nonconfinement in the nonrelativistic approach. The case $V_{v}=-V_{s}$ $+$ $%
\mathrm{constant}$ investigated in this work, for instance, presents bounded
solutions in the relativistic approach, although it reduces to the problem
of a particle subject to a uniform background potential in the
nonrelativistic limit. This last phenomenon is a consequence of the fact
that vector and scalar potentials couple differently in the relativistic
equations whereas there is no such distinction among them in the Schr\"{o}%
dinger equation. Regarding the structure of the wavefunctions under the
simultaneous changes $V_{v}\rightarrow-V_{v}$ and $E\rightarrow-E$, from the
charge-conjugation operation, one can see that if $\psi$ is a solution with
energy $E$ for the potential $V_{v}$, then $\sigma_{1}\psi^{\ast}$ is also a
solution with energy $-E$ for the potential $-V_{v}$. Thus, one has $%
(\psi_{\pm})_{c}=\psi^{\ast}_{\mp}$ and that means that the upper and lower
components of the Dirac spinor have their roles changed. As for the KG
wavefunction, its nodal structure is trivially preserved in such a way that
particle and antiparticle can be distinguished only by the eigenenergies.

Supposing that the vector and scalar potentials are constrained by the
relation $V_{v}+V_{s}=V_{0}$, where $V_{0}$ is a constant, and defining

\begin{equation}
m_{\mathtt{eff}}=|m+\frac{V_{0}}{c^{2}}|,\quad V_{0}\neq -mc^{2}  \label{8}
\end{equation}%
\begin{equation}
E_{\mathtt{eff}}=\frac{E^{2}-m_{\mathtt{eff}}^{2}c^{4}}{2m_{\mathtt{eff}%
}c^{2}}  \label{9}
\end{equation}%
\begin{equation}
V_{\mathtt{eff}}=\frac{E-m_{\mathtt{eff}}c^{2}\,\mathrm{sgn}\left(
mc^{2}+V_{0}\right) }{m_{\mathtt{eff}}c^{2}}\,V_{v}  \label{10}
\end{equation}%
the Klein-Gordon equation can be written as

\begin{equation}
-\frac{\hbar ^{2}}{2m_{\mathtt{eff}}}\,\phi ^{\prime \prime }+V_{\mathtt{eff}%
}\,\phi =E_{\mathtt{eff}}\,\phi  \label{11}
\end{equation}%
On the other hand, for $E\neq m_{\mathtt{eff}}c^{2}\,\mathrm{sgn}\left(
mc^{2}+V_{0}\right) $ the same Sturm-Liouville equation for $\phi $ is
obeyed by $\psi _{-}$ whereas%
\begin{equation}
\psi _{+}=-\frac{i\hbar c\psi _{-}^{\prime }}{E-m_{\mathtt{eff}}c^{2}\,%
\mathrm{sgn}\left( mc^{2}+V_{0}\right) }  \label{13}
\end{equation}%
Otherwise, for $E=m_{\mathtt{eff}}c^{2}\,\mathrm{sgn}\left(
mc^{2}+V_{0}\right) $, it might be possible the existence of an isolated
solution given by%
\begin{equation}
\psi _{-}=\mathrm{const.},\quad \psi _{+}=\frac{2i\psi _{-}}{\hbar c}%
\int^{x}dx\left( V_{s}+mc^{2}\right)  \label{14}
\end{equation}%
Of course, this solution does not exist if the domain is infinity because $%
\psi _{-}$ would not be square integrable. Note that apart from the possible
isolated solution, $\psi _{-}$ satisfies the KG equation. An equally
interesting result in the case of vanishing mass is that the spectrum just
changes sign when $V_{0}$ does. As for the eigenfunctions, $\phi $ and $\psi
_{-}$ are invariant under the change of the sign of $V_{0}$ whereas $\psi
_{+}$ changes sign.

For the specific case of the two-parameter potential functions $V_{v}=V_{0}\,%
\mathrm{sech}^2 \,\alpha x$ and $V_{s}=V_{0}\,\mathrm{tanh}^2 \,\alpha x$,
the isolated solution of the Dirac equation is not normalizable and the
effective potential of the Sturm-Liouville problem for both $\phi $ and $%
\psi _{-}$ can be expressed as%
\begin{equation}
V_{\mathtt{eff}}=-U_{0}\,\mathrm{sech}^{2}\,\alpha x,\quad U_{0}=\left[ m_{%
\mathtt{eff}}c^{2}\,\mathrm{sgn}\left( mc^{2}+V_{0}\right) -E\right] \frac{%
V_{0}}{m_{\mathtt{eff}}c^{2}}  \label{15}
\end{equation}%
Notice that $V_{\mathtt{eff}}$ is invariant under the change $\alpha
\rightarrow -\alpha $ so that the results can depend only on $|\alpha |$.
Furthermore, the effective potential is an even function under $x\rightarrow
-x$ in such way that $\phi $ and $\psi _{-}$ can be taken to be even or odd.
Note also that when $E>m_{\mathtt{eff}}c^{2}$ for $V_{0}>0$, $E<m_{\mathtt{%
eff}}c^{2}$ for $-mc^{2}<V_{0}<0$, and $E<-m_{\mathtt{eff}}c^{2}$ for $%
V_{0}<-mc^{2}$ one has $U_{0}<0$. In this case the effective potential is a
potential barrier \ and only scattering states are allowed with energies
restricted to $E>m_{\mathtt{eff}}c^{2}$ for $V_{0}>0$ and $E<-m_{\mathtt{eff}%
}c^{2}$ for $V_{0}<0$. Contrariwise, $U_{0}>0$ and the effective potential
is identified as the exactly solvable symmetric modified P\"{o}schl-Teller
potential (\cite{rm}, \cite{lan}-\cite{nie}). In this last case there is
also a continuum for $E<-m_{\mathtt{eff}}c^{2}$ for $V_{0}>0$ and $E>m_{%
\mathtt{eff}}c^{2}$ for $V_{0}<0$, and finite sets of discrete energies are
allowed in the ranges $-m_{\mathtt{eff}}c^{2}<E<2V_{0}-m_{\mathtt{eff}}c^{2}$
for $V_{0}>0$ and $|E|<m_{\mathtt{eff}}c^{2}$ for $V_{0}<-mc^{2}.$

Focusing attention on the bound-state solutions, one can see that the
normalizable eigenfunctions are subject to the boundary conditions $\phi
=\psi _{-}=0$ as $|x|\rightarrow \infty $ in such a manner that the solution
of our relativistic problem can be developed by taking advantage from the
knowledge of the exact solution for the symmetric modified P\"{o}schl-Teller
potential. The corresponding effective eigenenergy is given by ( \cite{rm},
\cite{lan}-\cite{nie})
\begin{equation}
E_{\mathtt{eff}}=-\,\frac{\hbar ^{2}\alpha ^{2}a_{n}^{2}}{2m_{\mathtt{eff}}}
\label{16}
\end{equation}

\noindent where%
\begin{equation}
a_{n}=s-n,\quad n=0,1,2,\ldots <s  \label{16a}
\end{equation}%
and

\begin{equation}
s=\frac{1}{2}\left( -1+\sqrt{1+\frac{8m_{\mathtt{eff}}U_{0}}{\hbar
^{2}\alpha ^{2}}}\,\right)   \label{17}
\end{equation}%
Note that the number of allowed bound states increases with $|V_{0}|$ and
decreases with $|\alpha |$, and that there is always at least one
bound-state solution. Now, equations (\ref{16})-(\ref{17}) lead to the
quantization condition%
\[
\sqrt{\hbar ^{2}c^{2}\alpha ^{2}+8V_{0}\left[ m_{\mathtt{eff}}c^{2}\,\mathrm{%
sgn}\left( mc^{2}+V_{0}\right) -E\right] }
\]%
\begin{equation}
-2\sqrt{m_{\mathtt{eff}}^{2}c^{4}-E^{2}}=\hbar c|\alpha |\left( 2n+1\right)
\label{19}
\end{equation}%
Once one makes sure one meets the requirement dictated by (\ref{16a}), \ the
solutions of (\ref{19}) determinate the eigenvalues of the relativistic
problem. This equation can be solved easily with a symbolic algebra program
by searching eigenenergies in the range $-m_{\mathtt{eff}}c^{2}<E<2V_{0}-m_{%
\mathtt{eff}}c^{2}$ for $V_{0}>0$ and $|E|<m_{\mathtt{eff}}c^{2}$ for $%
V_{0}<-mc^{2}$, as foreseen by the preceding qualitative arguments. Of
course, for $V_{0}>0$ one obtains $E\approx -mc^{2}$ when $V_{0}\ll mc^{2}$
and $-V_{0}<E<V_{0}$ when $V_{0}\gg mc^{2}$. One the other hand, for $%
V_{0}<-mc^{2}$ one finds $E\approx 0$ when $V_{0}\approx -mc^{2}$, and $%
-|V_{0}|<E<|V_{0}|$ when $|V_{0}|\gg mc^{2}$.

It happens that there is at most one solution of (\ref{19}) for a given
quantum number. Do these energies are related to particle or antiparticle
energy levels? To answer this question we plot the energy levels in terms of
the parameters of the potential. Figures 1 and 2 show the behaviour of the
energies as a function of \ $V_{0}$ and $\alpha $, respectively. From Fig. 1
one sees that all the energy levels emerge from the negative-energy
continuum so that it is plausible to identify them with antiparticle levels,
although for a given $V_{0}$ some of the levels can have positive energies.
Meanwhile, from Fig. 2 one sees that the energy levels tend to disappear one
after another as $\alpha $ increases and just the ground-state energy level
survives as $\alpha \rightarrow \infty $. The energy levels passing out of
the picture as $\alpha $ increases (or $V_{0}$ decreases as in Fig. 1) sink
into the negative continuum but this does not menace the single-particle
interpretation of either KG or Dirac equations since one has antiparticle
levels plunging into the antiparticle continuum. It is also noticeable from
both of these figures that for a given set of potential parameters one finds
that the lowest quantum numbers correspond to the highest eigenenergies, as
it should be for antiparticle energy levels. For $V_{0}<-mc^{2}$ the
spectrum presents a similar behaviour but the energy levels emerge from the
upper continuum and are to be identified with particle levels. If we had
plotted the spectra for a massless particle, we would encounter, up to the
sign of $E$, identical spectra for both signs of $V_{0}$. At any
circumstance, the spectrum contains either particle-energy levels or
antiparticle-energy levels. This conclusion confirms what has already been
analyzed in \cite{beni}: the spectrum contains either particle-energy levels
or antiparticle-energy levels depending on the sign of the coupling constant.

The KG eigenfunction as well as the lower component of the Dirac spinor can
be given by (\cite{nie})%
\[
\phi =\psi _{-}=N\,2^{a_{n}}\Gamma \left( a_{n}+\frac{1}{2}\right) \sqrt{%
\frac{|\alpha |a_{n}}{\pi }\frac{\Gamma \left( n+1\right) }{\Gamma \left(
n+1+2a_{n}\right) }}
\]

\begin{equation}
\,\times \left( 1-z^{2}\right) ^{a_{n}/2}C_{n}^{\left( a_{n}+1/2\right)
}\left( z\right)  \label{21}
\end{equation}

\noindent where $z=\mathrm{tanh}\,\alpha x$ and $C_{n}^{\left( a\right)
}\left( z\right) $ is the Gegenbauer (ultraspherical) polynomial of degree $%
n $. Since $C_{n}^{\left( a\right) }\left( -z\right) =\left( -\right)
^{n}C_{n}^{\left( a\right) }\left( z\right) $ and $C_{n}^{\left( a\right)
}\left( z\right) $ has $n$ distinct zeros (see, e.g. \cite{abr}), it becomes
clear that $\psi _{+}$ and $\psi _{-}$ have definite and opposite parities.
The constant $N$ \ is the unit in the KG problem and it chosen such that $%
\int_{-\infty }^{+\infty }dx\left( |\psi _{+}|^{2}+|\psi _{-}|^{2}\right) =1$
in the Dirac problem. Fig. 3 illustrates the behavior of the upper and lower
components of the Dirac spinor $|\psi _{+}|^{2}$ and $|\psi _{-}|^{2}$, and
the position probability densities $|\psi |^{2}=|\psi _{+}|^{2}+|\psi
_{-}|^{2}$ and $|\phi |^{2}$ for $n=0$. The relative normalization constant
was calculated numerically. Comparison of $|\psi _{+}|^{2}$ and $|\psi
_{-}|^{2}$ shows that $\psi _{+}$ $\ $\ is suppressed relative to $\psi _{-}$%
. This result is expected since we have here an antiparticle eigenstate.
Nevertheless, the same behavior shows its face for the particle eigenstates
(for $V_{0}<-mc^{2}$). One might say that this kind of effect is because we
are dealing with a quintessential relativistic potential. In addition,
comparison of $\ |\phi |^{2}|$ and $|\psi |^{2}$ shows that a KG particle
tends to be better localized than a Dirac particle. As a matter of fact, a
numerical calculation of the uncertainty in the position (with $m=c=\hbar =1$
and $V_{0}=\alpha =5$) furnishes $0.160$ and $0.179$, respectively. Here we
have purposely shown an odd fact. It seems that the uncertainty principle
dies away provided such a principle implies that it is impossible to
localize a particle into a region of space less than half of its Compton
wavelength (see, e.g., Ref. \cite{str}). This apparent contradiction can be
remedied by recurring to the concept of effective Compton wavelength, as has
been done previously in connection with pseudoscalar couplings in the Dirac
equation \cite{asc}. Indeed, Eq. (\ref{8}) suggests that we can define the
effective Compton wavelength as $\lambda _{\mathtt{eff}}=\hbar /(m_{\mathtt{%
eff}}c)$ so that the minimum uncertainty consonant with the uncertainty
principle is given by $\lambda _{\mathtt{eff}}/2$ whereas the maximum
uncertainty in the momentum is given by $m_{\mathtt{eff}}c$. The
appropriateness of the concept of effective Compton wavelength has been
checked for a large range of the potential parameters.

In summary, the methodology for finding solutions of the KG and the Dirac
equations for the enlarged class of mixed vector-scalar potentials
satisfying the constraint $V_{v}+V_{s}= V_{0}$ have been put forward.
Although the KG and the Dirac equations exhibit the very same spectrum their
eigenfunctions make all the difference. With the two-parameter potential
functions $V_{v}=V_{0}\,\mathrm{sech}^2 \,\alpha x$ and $V_{s}=V_{0}\,%
\mathrm{tanh}^2 \,\alpha x$, the KG and the Dirac equations have been mapped
into a Schr\"{o}dinger-like equation with the symmetric modified P\"{o}%
schl-Teller potential and we have shown that a KG particle tends to be
better localized than a Dirac particle. In both cases, the spectrum consists
of either particles or antiparticles, depending on the sign of $V_{0}$. An
apparent contradiction with the uncertainty principle was cured by
introducing the effective Compton wavelength.

\newpage

\begin{figure}[th]
\begin{center}
\includegraphics[width=9cm, angle=270]{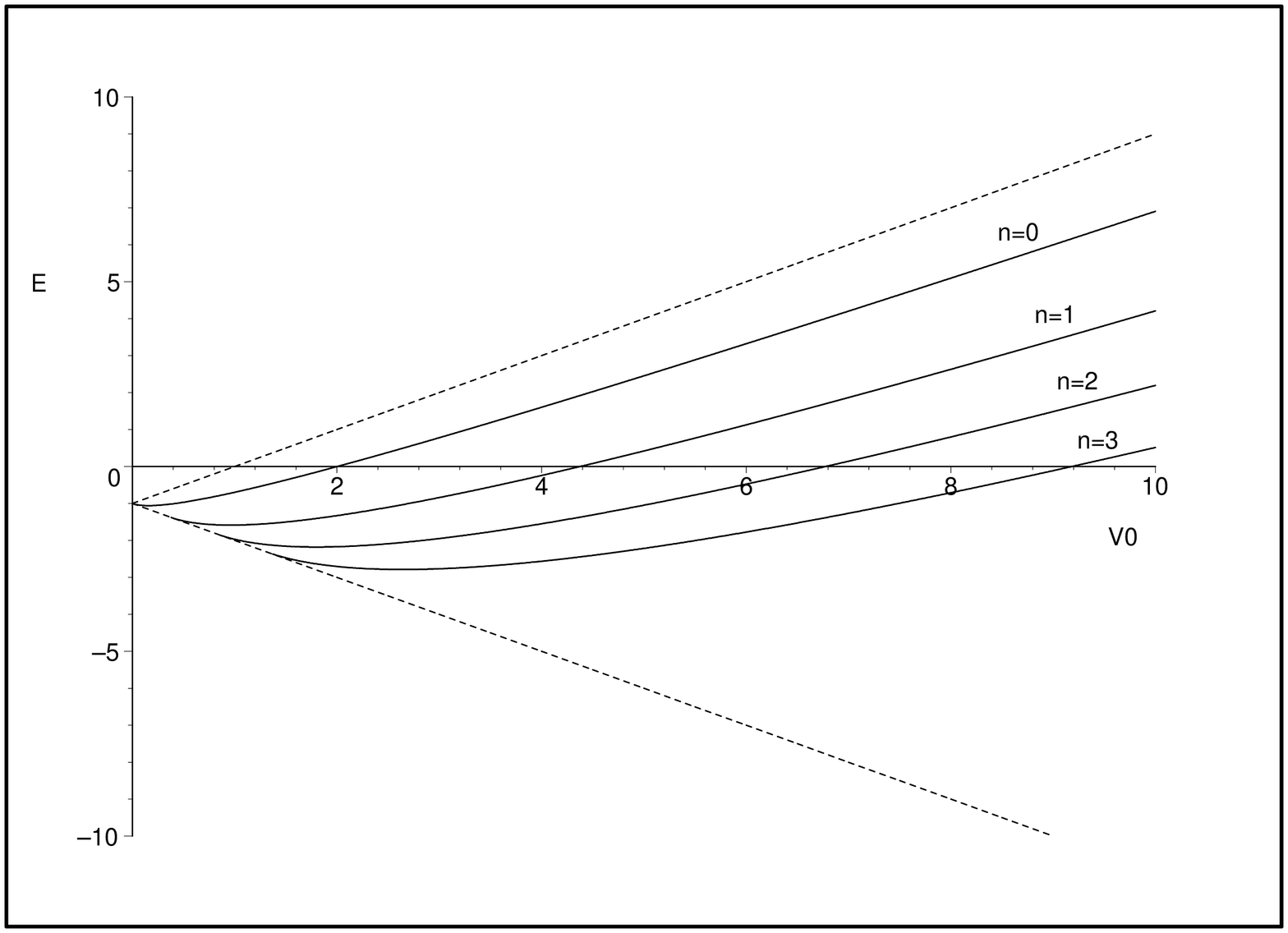}
\end{center}
\par
\vspace*{-0.1cm}
\caption{Dirac eigenvalues for the four lowest quantum numbers as a function
of $V_{0}$. The dashed lines stand for $-m_{\mathtt{eff}}c^{2}$ and $%
2V_{0}-m_{\mathtt{eff}}c^{2}$ ($m=\hbar=c=\protect\alpha=1$). }
\label{Fig1}
\end{figure}

\begin{figure}[th]
\begin{center}
\includegraphics[width=9cm, angle=270]{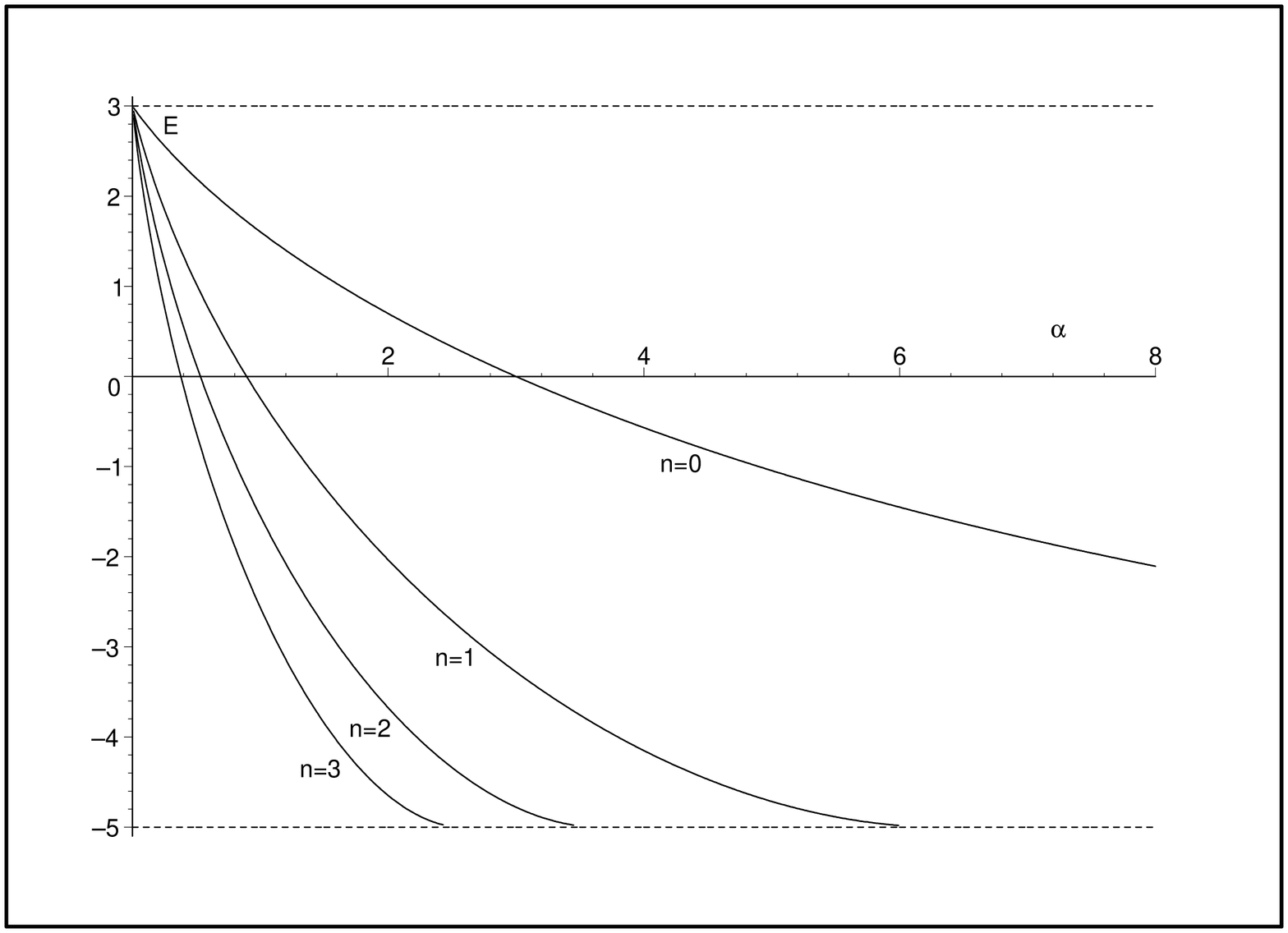}
\end{center}
\par
\vspace*{-0.1cm}
\caption{Dirac eigenvalues for the four lowest quantum numbers as a function
of $\protect\alpha$. The dashed lines stand for $-m_{\mathtt{eff}}c^{2}$ and
$2V_{0}-m_{\mathtt{eff}}c^{2}$ ($m=\hbar=c=1$ and $V_{0}=4$). }
\label{Fig2}
\end{figure}

\begin{figure}[th]
\begin{center}
\includegraphics[width=9cm, angle=270]{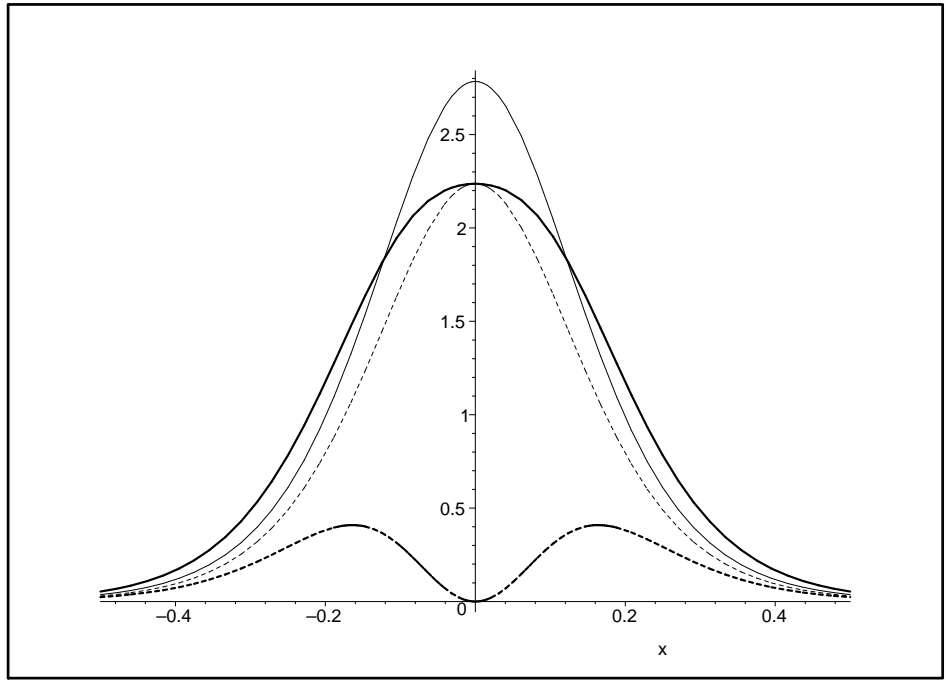}
\end{center}
\par
\vspace*{-0.1cm}
\caption{$|\protect\psi _{+}|^{2}$ (heavy dashed line), $|\protect\psi %
_{-}|^{2}$ (light dashed line), $|\protect\psi |^{2}=|\protect\psi %
_{+}|^{2}+|\protect\psi _{-}|^{2}$ (thick line) and $|\protect\phi |^{2}$
(thin line) for $n=0$ ($m=\hbar=c=1$ and $V_{0}=\protect\alpha = 5$). }
\label{Fig3}
\end{figure}

\end{document}